\documentclass[conference]{IEEEtran}
\IEEEoverridecommandlockouts
\usepackage{cite}
\usepackage{amsmath,amssymb,amsfonts}
\usepackage{algorithmic}
\usepackage{graphicx}
\usepackage{textcomp}
\usepackage{xcolor}
\usepackage{xspace}
\usepackage{booktabs}
\usepackage{tabularray}
\UseTblrLibrary{booktabs}
\usepackage{xurl}
\usepackage{seqsplit}
\newcommand{\rqcode}[1]{{\ttfamily\tiny\seqsplit{#1}}}
\newcommand{\rqfn}[1]{{\ttfamily\seqsplit{#1}}}      
\newcommand{\rqfnshort}[1]{{\ttfamily #1}}           
\usepackage{pifont}
\usepackage{hyperref}

\usepackage{tikz}
\usetikzlibrary{positioning,arrows,arrows.meta,fit,backgrounds,calc}

\definecolor{warncol}{HTML}{C47A1A}
\newsavebox{\bombiconbox}
\savebox{\bombiconbox}{%
  \begin{tikzpicture}[baseline=-1pt, scale=0.11]
    \fill[black!80] (0,0) circle (1.15);
    \draw[black!80, line width=1.1pt, rounded corners=0.8pt]
      (0.1,1.1) .. controls (0.45,1.75) and (0.85,1.95) .. (1.05,2.35);
    \fill[warncol] (1.05,2.35) circle (0.22);
  \end{tikzpicture}%
}

\usepackage{listings}
\usepackage{xcolor}
\usepackage{relsize} 

\definecolor{verylightgray}{rgb}{.97,.97,.97}
\definecolor{lightyellow}{rgb}{1.0,0.97,0.75}
\definecolor{lightgreen}{rgb}{0.839,0.961,0.839}
\definecolor{lightred}{rgb}{1.0,0.85,0.85}

\definecolor{darkgreen}{rgb}{0.0, 0.5, 0.0}

\newcommand{\cmark}{\textcolor{darkgreen}{\ding{51}}}%

\lstdefinelanguage{Solidity}{
	keywords=[1]{anonymous, assembly, assert, balance, break, call, callcode, case, catch, class, constant, continue, constructor, contract, debugger, default, delegatecall, delete, do, else, emit, event, experimental, export, external, false, finally, for, function,  gas, if, implements, import, in, indexed, instanceof, interface, internal, is, length, library, log0, log1, log2, log3, log4, memory, modifier, new, payable, pragma, private, protected, public, pure, push, require, return, returns, revert, selfdestruct, send, solidity, storage, struct, suicide, super, switch, then, this, throw, transfer, true, try, typeof, using, value, view, while, with, addmod, ecrecover, keccak256, mulmod, ripemd160, sha256, sha3}, 
	keywordstyle=[1]\color{blue}\bfseries,
	keywords=[2]{address, bool, byte, bytes, bytes1, bytes2, bytes3, bytes4, bytes5, bytes6, bytes7, bytes8, bytes9, bytes10, bytes11, bytes12, bytes13, bytes14, bytes15, bytes16, bytes17, bytes18, bytes19, bytes20, bytes21, bytes22, bytes23, bytes24, bytes25, bytes26, bytes27, bytes28, bytes29, bytes30, bytes31, bytes32, enum, int, int8, int16, int24, int32, int40, int48, int56, int64, int72, int80, int88, int96, int104, int112, int120, int128, int136, int144, int152, int160, int168, int176, int184, int192, int200, int208, int216, int224, int232, int240, int248, int256, mapping, string, uint, uint8, uint16, uint24, uint32, uint40, uint48, uint56, uint64, uint72, uint80, uint88, uint96, uint104, uint112, uint120, uint128, uint136, uint144, uint152, uint160, uint168, uint176, uint184, uint192, uint200, uint208, uint216, uint224, uint232, uint240, uint248, uint256, var, void, ether, finney, szabo, wei, days, hours, minutes, seconds, weeks, years},	
	keywordstyle=[2]\color{teal}\bfseries,
	keywords=[3]{block, blockhash, coinbase, difficulty, gaslimit, number, timestamp, msg, data, gas, sender, sig, value, now, tx, gasprice, origin},	
	keywordstyle=[3]\color{violet}\bfseries,
	identifierstyle=\color{black},
	sensitive=true,
	comment=[l]{//},
	morecomment=[s]{/*}{*/},
	commentstyle=\color{gray}\ttfamily,
	stringstyle=\color{red}\ttfamily,
	morestring=[b]',
	morestring=[b]"
}

\newcommand{\lsthl}[2]{%
  \colorbox{#1}{%
    \begin{minipage}[t]{\linewidth}%
      \strut\ttfamily\footnotesize #2%
    \end{minipage}%
  }%
}
\newcommand{\invariantmark}[1]{\lsthl{lightyellow}{#1}}
\newcommand{\diffaddmark}[1]{\lsthl{lightgreen}{#1}}

\lstdefinestyle{invariant}{
	language=Solidity,
	backgroundcolor=\color{verylightgray},
	extendedchars=true,
	basicstyle=\footnotesize\ttfamily,
	showstringspaces=false,
	showspaces=false,
	numbers=left,
	numberstyle=\footnotesize\ttfamily,
	numbersep=3pt,
	tabsize=1,
	breaklines=true,
	showtabs=false,
	captionpos=b,
	xleftmargin=0.5em,
	framexleftmargin=1.5em,
	xrightmargin=1em,
	escapeinside={@}{@},
}

\lstdefinestyle{diff}{
	language=Solidity,
	backgroundcolor=\color{verylightgray},
	extendedchars=true,
	basicstyle=\footnotesize\ttfamily,
	showstringspaces=false,
	showspaces=false,
	numbers=left,
	numberstyle=\footnotesize\ttfamily,
	numbersep=3pt,
	tabsize=1,
	breaklines=true,
	showtabs=false,
	captionpos=b,
	xleftmargin=0.5em,
	framexleftmargin=1.5em,
	xrightmargin=1em,
	escapeinside={@}{@},
}

\usepackage{xcolor}
\definecolor{eclipseKeywords}{RGB}{127,0,85}

\usepackage[export]{adjustbox}
\usepackage{tcolorbox} 

\newtcolorbox{answerbox}[1]{
    colback=gray!10!white,
    colframe=black!40!white,
    coltitle=black,
    title={#1},
    fonttitle=\bfseries,
    attach title to upper,
    after title={:\ },
    boxrule=0.8pt,
    width=0.96\linewidth,
    center
}
\definecolor{codegray}{rgb}{0.95,0.95,0.95}

\newcommand{\pondereplay}{\textsc{PonDeReplay}\xspace}

\newcommand{\defihacklabs}{\textsc{DeFiHackLabs}\xspace}
\newcommand{\aegis}{\text{\AE}GIS\xspace}
\newcommand{\dataset}{\textsc{InvariantEval}\xspace}
\newcommand{\datasetsize}{$28$\xspace}
\newcommand{\dfhltotal}{$871$\xspace}       
\newcommand{\ethhacks}{$359$\xspace}         
\newcommand{\rqone}{To what extent do program invariants stop real world attacks on smart contracts?\xspace}

\newcommand{\rqtwo}{To what extent do attack-stopping invariants break existing behavior according to historical replay verification?\xspace}

\newcommand{\rqthree}{What are the limitations of historical replay verification?\xspace}

\newcommand{\rqfour}{To what extent are the state-of-the-art invariant generation tools able to generate attack-stopping invariants for real world attacks?\xspace}

\newcommand{\totalnumreplayed}{$108{,}637$\xspace}

\def\BibTeX{{\rm B\kern-.05em{\sc i\kern-.025em b}\kern-.08em
    T\kern-.1667em\lower.7ex\hbox{E}\kern-.125emX}}
\begin{document}

\title{Smart Contract Invariants Protect Against Cybercriminals}

\author{
\IEEEauthorblockN{
Sofia Bobadilla,
Humaira Afrin,
Angela Novelli,
Martin Monperrus}
\IEEEauthorblockA{
KTH Royal Institute of Technology\\
Stockholm, Sweden\\
\{sofbob, humairaa, anovelli, monperrus\}@kth.se
}
}

\maketitle
\begin{abstract}
Blockchains are among the most adversarial environments in computing.
Billions are stolen by cybercriminals who exploit vulnerabilities.
This is an open problem and no concept or technique has proven to really make a difference.
In this paper, we claim that the classical notion of program invariant is perhaps the most powerful solution to the problem.
We devise an original experimental protocol to 1) study how invariants would have protected against past real-world attacks and 2) whether state-of-the-art automated tools can find them.
The experimental toolchain is sophisticated.
It is based on \dataset, a benchmark of \datasetsize real Ethereum exploits, each paired with a human-authored invariant that blocks the attack.
We validate every invariant with \pondereplay, a replay framework that re-executes transactions in order to prove the correctness and soundness of smart contract invariants.
We demonstrate that smart contract invariants block all the cybercriminal attacks in \dataset, fully validated by replaying \totalnumreplayed historical transactions.
Our large-scale experiments clearly demonstrate that smart contract invariants protect against cybercriminals.
\end{abstract}

\section{Introduction}

Smart contract exploits cause large and irreversible financial losses.
More than one billion dollars in on-chain assets have been stolen through vulnerabilities in deployed contracts~\cite{defihacklabs}.
A missing access-control guard lets any caller drain funds (e.g., Cork Protocol, \$12M~\footnote{\url{https://dedaub.com/blog/the-11m-cork-protocol-hack-a-critical-lesson-in-uniswap-v4-hook-security/}}).
A missing oracle check opens the door to price manipulation (e.g., Makina, \$5.1M~\footnote{\url{https://github.com/SunWeb3Sec/DeFiHackLabs/blob/main/src/test/2026-01/makina\_exp.sol}}).

Smart contract invariants are a powerful defense mechanism.
Developers write such invariants, and they are evaluated at runtime for all transactions. The blockchain guarantees are strong: if the invariant is violated, the transaction is reverted, and no side-effect or state change is committed to the chain.

Two important questions follow. Could a program invariant have stopped real-world exploits?
Can automated invariant generation tools discover those attack-stopping invariants?
While important, these are open, unanswered research questions.
Prior work evaluates invariant-related tools only against synthetic properties, templates, or mined patterns~\cite{liu_invcon_2022,chen_demystifying_2024,wang_smartinv_2024}, but none have been confronted with real-world, multi-million dollar loss incidents.

In this paper, we answer both questions with a novel protocol and a large-scale experiment. 
First, we build \dataset, a benchmark of \datasetsize real-world Ethereum exploits, each paired with a human-authored program invariant that blocks the attack.
Next, we validate each program invariant with a novel historical verification technique consisting of re-executing 1) the proof-of-concept exploit, 2) the attack transaction, and 3) the contract's real transaction history.
The replay enables us to verify security and functional preservation with guarantees.

The hardened contracts block all \datasetsize attacks: every ground-truth invariant causes both the proof-of-concept exploit and the malicious on-chain transaction to fail at the vulnerable step, a $28/28$ result.
We replay \totalnumreplayed historical transactions; of those, $106{,}815$ ($98.3\%$ of total replayed) are fully preserved under the hardened bytecode; the $1{,}526$ divergences ($1.4\%$) are each causally attributed to the invariant, not to replay noise.
The attack-stopping invariants reduce to eight recurring property families, with price and oracle checks ($8$ cases), ledger conservation ($6$), and access control and interaction safety ($4$ each) as the largest groups.

Finally, we study whether the state-of-the-art tools for invariant generation protect against real-world attacks. The publicly available, end-to-end executable tools fall short: FLAMES \cite{eshghie_flames_2025}, InvCon \cite{liu_invcon_2022}, and InvCon+ \cite{invconplus} together recover only 2 of the \datasetsize attack-stopping invariants.

Our contributions are:
\begin{itemize}
    \item an original, sound methodology for studying smart contract invariant effectiveness, fully grounded on on-chain data.
    \item \dataset\footnote{At the moment, access to \dataset and \pondereplay is available upon request to sofbob@kth.se.\label{fn:request}}, a benchmark of \datasetsize real Ethereum exploits each paired with a human-authored, attack-stopping program invariant
    \item \pondereplay~\footref{fn:request}, an on-chain verification framework to test an invariant against both the attack transaction and the contract's full pre-attack history, providing causal attribution to the protective invariant. Our experimental conclusions are grounded in \totalnumreplayed replays.
    \item An evaluation of three state-of-the-art invariant tools against the \dataset ground truth, showing the fundamental shortcomings of the field. 
\end{itemize}

The remainder of the paper is organized as follows. Section~\ref{sec:background} introduces invariants and smart contracts. Section~\ref{sec:methodology} describes the experimental methodology, the research questions, and the \pondereplay replay infrastructure. Section~\ref{sec:results} reports the results, and Section~\ref{sec:related} discusses related work.

\section{Background}
\label{sec:background}

\subsection{Blockchain Concepts}
\label{sec:blockchain-concepts}

\textbf{Smart Contract \& Solidity.}
Smart Contracts are programs deployed on blockchain networks. 
Contracts are written in high-level smart contract languages, such as Solidity, and then compiled into bytecode for deployment and execution. 
Solidity is Ethereum’s primary smart contract programming language. 
The Ethereum Virtual Machine (EVM) is the runtime environment for executing smart contracts on the Ethereum blockchain.

\textbf{Transactions.}
All operations on the blockchain are initiated via transactions. 
A transaction is typically sent from a user, signed by their private key, to perform actions such as transferring cryptocurrency or invoking functions on existing contracts.
During execution, a called contract may interact with other contracts. Transactions are atomic. If they revert, no side-effect is written on chain.

\textbf{Smart Contract Security.}
Before deployment, contracts are typically reviewed through a security audit: a manual and tool-assisted inspection by specialized firms that looks for vulnerabilities in the code~\cite{poco}.
Audits reduce risk but do not eliminate it.
Consequently, over one billion dollars in assets have been lost to smart contract exploits~\cite{defillama}.

\textbf{Proof-of-Concept.} Once an incident occurs, security researchers reconstruct the attack as a proof-of-concept (PoC) exploit.
A proof-of-concept is executable code that reproduces the theft on a fork of the blockchain at the vulnerable block.
DeFiHackLabs~\cite{defihacklabs} maintains a public collection of such PoCs for documented exploits.

\subsection{Invariants}
\label{sec:invariants}

An invariant is a property that holds at a designated point in a program~\cite{daikon2007}.
Three canonical forms are widely recognised:

\begin{itemize}
    \item \textbf{Preconditions} hold at function entry and constrain the inputs or state under which a function may be lawfully called (Listing~\ref{lst:pre}).
    \item \textbf{Postconditions} hold at function exit and capture guarantees about the output or resulting state (Listing~\ref{lst:post}).
    \item \textbf{Object invariants} (in OOP) hold over the entire lifetime of an object, capturing structural properties of the object's state that must never be violated. A smart contract deployed on chain is conceptually equivalent to an object (Listing~\ref{lst:object}).
\end{itemize}

\begin{lstlisting}[style=invariant, label={lst:pre}, caption={Precondition invariant: input bound before batch transfer (BEC)},float]
function batchTransfer(
    address[] _receivers, uint256 _value
) public whenNotPaused returns (bool) {
    uint256 cnt = _receivers.length;
@\invariantmark{require(\_value <= uint256(-1) / cnt);}@
    // ... transfer loop ...
}
\end{lstlisting}

\begin{lstlisting}[style=invariant, label={lst:post}, caption={Postcondition invariant: output bound after Curve exchange (ERC20TokenBank)},float]
function doExchange(uint256 amount) public returns (bool) {
    // ... pull tokens, call Curve ...
    curve.exchange_underlying(1, 2, camount, 0);
@\invariantmark{require(namount >= (camount * 995) / 1000, "slippage too high");}@
}
\end{lstlisting}

\begin{lstlisting}[style=invariant, label={lst:object}, caption={Object invariant: a reentrancy lock, held across every state-mutating method via a modifier},float]
modifier nonReentrant() \{
@\invariantmark{require(!locked, "reentered");}@
    locked = true; _; locked = false;

@\invariantmark{function buyJay(...) public payable nonReentrant}@ { /* ... */ }
\end{lstlisting}

Invariants appear across the software engineering literature under many names: \emph{assertions}, \emph{contracts}, \emph{properties}, \emph{runtime checks}.
They all express the same idea: a logical condition that must hold at a designated program point.
Developers rarely write them well~\cite{daikon2007}.

\textbf{Invariants for Security}
A security invariant serves a strong purpose: it encodes a property that must hold even under adversarial input.
When such an invariant is checked at runtime, a violation basically means that the attack is blocked and the offending execution is reverted before it commits any effect.
Invariants for smart contract security is a promising direction, with active research in the area, see \autoref{sec:related}.

\section{Experimental Methodology}
\label{sec:methodology}
We study the extent to which program invariants can stop real exploits, how much those invariants modify behavior, and how well current tools can discover them.
This research is essential for understanding the current capabilities and steering future research.

\subsection{Overview}\label{sec:overview}

Our experimental methodology answers two main questions: can a single program invariant stop each real exploit, and can automated tools discover it?

We build on \defihacklabs, a corpus of executable real-world exploits against real on-chain contracts.
For each selected exploit, two authors independently write a \emph{hardened contract}: the original source code augmented with a program invariant inserted at the vulnerable function.

Each invariant is then validated by \pondereplay, a novel replay framework we built for this purpose.
\pondereplay re-executes two classes of transactions against the hardened bytecode: the attack transactions (to confirm the invariant stops the exploit) and historical transactions (to confirm normal usage is preserved).
An invariant is accepted only when it blocks the attack and leaves legitimate behavior intact or with explained changes.

This design grounds every claim we will make in on-chain evidence: real calldata, real block state, and real execution outcomes.
Figure~\ref{fig:verification-pipeline} summarizes the verification pipeline.
Section~\ref{sec:dataset} describes our dataset of real-world attacks.
Section~\ref{sec:writing} describes the invariant authoring process, Section~\ref{sec:verification} the joint security and preservation checks, and Section~\ref{sec:pondereplay} the \pondereplay infrastructure.
The four research questions are then evaluated against this validated ground truth.


\definecolor{passcol}{HTML}{2E7D52}
\definecolor{passfill}{HTML}{E9F4EE}
\definecolor{failcol}{HTML}{B0413E}
\definecolor{failfill}{HTML}{FBECEB}
\definecolor{inkgray}{HTML}{333333}
\definecolor{secfill}{HTML}{FBECEB}   
\definecolor{secedge}{HTML}{B0413E}
\definecolor{prsfill}{HTML}{EAF1FA}   
\definecolor{prsedge}{HTML}{3B5C8F}

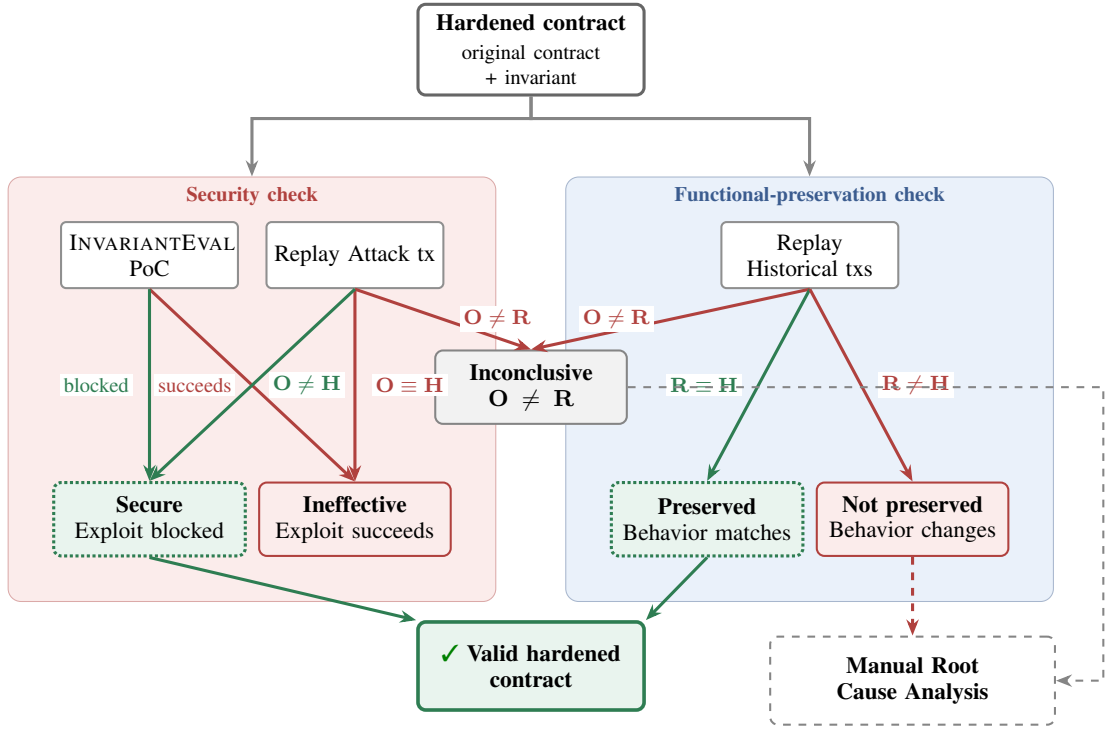
\begin{figure*}[t]
  \centering
  \resizebox{.8\linewidth}{!}{%
  \begin{tikzpicture}[
    font=\small,
    hardened/.style = {rectangle, rounded corners=3pt, draw=inkgray!75, fill=white, very thick,
                       align=center, inner sep=4pt, minimum height=12mm, text width=28mm},
    box/.style     = {rectangle, rounded corners=2pt, draw=inkgray!55, fill=white, thick,
                       align=center, inner sep=3pt, minimum height=9mm, text width=22mm},
    good/.style    = {rectangle, rounded corners=3pt, draw=passcol, densely dotted, fill=passfill,
                       line width=1.2pt, align=center, inner sep=3pt, minimum height=10mm, text width=24mm},
    bad/.style     = {rectangle, rounded corners=3pt, draw=failcol, fill=failfill, thick,
                       align=center, inner sep=3pt, minimum height=10mm, text width=24mm},
    inconc/.style  = {rectangle, rounded corners=3pt, draw=inkgray!60, fill=gray!10, thick,
                       align=center, inner sep=3pt, minimum height=10mm, text width=24mm},
    valid/.style   = {rectangle, rounded corners=3pt, draw=passcol, fill=passfill, line width=1.4pt,
                       align=center, inner sep=4pt, minimum height=12mm, text width=28mm},
    manual/.style  = {rectangle, rounded corners=3pt, draw=inkgray!60, dashed, thick,
                       align=center, inner sep=4pt, minimum height=12mm, text width=36mm},
    ttl/.style     = {font=\bfseries\footnotesize},
    arr/.style     = {-{Stealth[length=2.2mm,width=1.7mm]}, very thick, draw=inkgray!60},
    okarr/.style   = {-{Stealth[length=2.2mm,width=1.7mm]}, very thick, draw=passcol},
    badarr/.style  = {-{Stealth[length=2.2mm,width=1.7mm]}, very thick, draw=failcol},
    darr/.style    = {-{Stealth[length=2.2mm,width=1.7mm]}, thick, draw=inkgray!60, dashed},
    lbl/.style     = {font=\footnotesize, fill=white, inner sep=1.4pt}
  ]

  \node[hardened] at (0,58mm) (hard)
    {{\textbf{Hardened contract}}\\[1pt]\footnotesize original contract + invariant};

  \node[ttl, text=secedge] at (-38mm,38mm) (seclabel) {Security check};
  \node[ttl, text=prsedge] at (38mm,38mm) (prslabel) {Functional-preservation check};

  \node[box] at (-52mm,30mm) (poc) {\dataset{} PoC};
  \node[box] at (-24mm,30mm) (attack) {Replay Attack tx};
  \node[box] at (38mm,30mm) (benign) {Replay Historical txs};

  \node[inconc] at (0,12mm) (inconc)
    {\textbf{Inconclusive}\\ $\mathbf{O} \neq \mathbf{R}$};

  \node[good] at (-52mm,-6mm) (secure)
    {\textbf{Secure}\\ Exploit blocked};
  \node[bad] at (-24mm,-6mm) (ineff)
    {\textbf{Ineffective}\\ Exploit succeeds};
  \node[good] at (24mm,-6mm) (preserved)
    {\textbf{Preserved}\\ Behavior matches};
  \node[bad] at (52mm,-6mm) (notpres)
    {\textbf{Not preserved}\\ Behavior changes};

  \node[valid] at (0,-26mm) (valid)
    {{\cmark}\ \textbf{Valid hardened contract}};

  \node[manual] at (52mm,-28mm) (manual)
    {\textbf{Manual Root Cause Analysis}};

  \begin{scope}[on background layer]
    \node[fit=(poc)(attack)(secure)(ineff), draw=secedge!45, fill=secfill,
          rounded corners=5pt, inner sep=6mm] (secbox) {};
    \node[fit=(benign)(preserved)(notpres), draw=prsedge!45, fill=prsfill,
          rounded corners=5pt, inner sep=6mm] (presbox) {};
  \end{scope}

  \draw[arr] (hard.south) -- ++(0,-3mm) -| (secbox.north);
  \draw[arr] (hard.south) -- ++(0,-3mm) -| (presbox.north);

  \draw[badarr] (poc.south) -- (ineff.north)
    node[lbl, pos=0.5, left=6pt, text=failcol] {succeeds};
  \draw[okarr]  (poc.south) -- (secure.north)
    node[lbl, pos=0.5, left=6pt, text=passcol] {blocked};

  \draw[badarr] (attack.south) -- (inconc.north)
    node[lbl, pos=0.5, right=6pt, text=failcol] {$\mathbf{O} \neq \mathbf{R}$};

  \draw[badarr] (attack.south) -- (ineff.north)
    node[lbl, pos=0.5, right=6pt, text=failcol] {$\mathbf{O} \equiv \mathbf{H}$};
  \draw[okarr]  (attack.south) -- (secure.north)
    node[lbl, pos=0.5, right=6pt, text=passcol] {$\mathbf{O} \neq \mathbf{H}$};

  \draw[badarr] (benign.south) -- (inconc.north)
    node[lbl, pos=0.5, left=6pt, text=failcol] {$\mathbf{O} \neq \mathbf{R}$};

  \draw[okarr]  (benign.south) -- (preserved.north)
    node[lbl, pos=0.5, left=6pt, text=passcol] {$\mathbf{R} \equiv \mathbf{H}$};
  \draw[badarr] (benign.south) -- (notpres.north)
    node[lbl, pos=0.5, right=6pt, text=failcol] {$\mathbf{R} \neq \mathbf{H}$};

  \draw[okarr] (secure.south) -- (valid.north west);
  \draw[okarr] (preserved.south) -- (valid.north east);

  \draw[badarr, dashed] (notpres.south) -- (manual.north);

  \draw[darr] (inconc.east) -- (78mm,12mm) -- (78mm,-28mm) -- (manual.east);

  \end{tikzpicture}%
  }
  \caption{Historical replay verification of a hardened contract.
  Each transaction yields three traces: the on-chain trace ($\mathbf{O}$), the recompiled trace of the unmodified contract ($\mathbf{R}$), and the hardened-contract trace ($\mathbf{H}$); $\equiv$ denotes trace equivalence per our fidelity criteria (Section~\ref{sec:replay-fidelity}).
  The \textbf{security check} replays the proof-of-concept exploit and the malicious transaction; the \textbf{functional-preservation check} replays historical transactions through the hardened function.}
  \label{fig:verification-pipeline}
\end{figure*}

\subsection{Dataset}\label{sec:dataset}

First, we need a dataset of real world attacks, in order to study to what extent invariants can actually protect smart contracts. We do not use toy examples like SmartBugs-Curated~\cite{smartbugs} or synthetic contracts like Iuliano et al.~\cite{injectedvuln}.

Our idea is to build our study on the well established \defihacklabs benchmark~\cite{defihacklabs}.
\defihacklabs is an open-source repository of real-world DeFi proof-of-concept exploits (PoCs) implemented in Solidity using the Foundry testing framework.
Each exploit comes with a self-contained, executable test that is meant to reproduce a documented past attack against its original on-chain contracts.
The full repository contains \dfhltotal attacks across multiple blockchains, including Ethereum, BNB Chain, and Polygon.

We restrict our study to Ethereum, arguably the largest smart contract platform by total money locked and home base to the majority of high-profile DeFi protocols, such as Uniswap and AAVE.
Restricting to one chain keeps our experimental environment uniform. This leaves \ethhacks Ethereum exploits, spanning July 2017 to June 2026.

\subsection{Ground Truth: Writing Program Invariants}
\label{sec:writing}

Prior invariant studies evaluate smart contract invariant effectiveness with predefined properties, templates, or mined patterns~\cite{liu_invcon_2022,chen_demystifying_2024,wang_smartinv_2024}.
Our core novelty is to entirely ground our study of smart contract invariant effectiveness in real-world attacks.

Our intuition is that the interesting invariants are \emph{attack-stopping}. If one has had them in the code, the attack would have never succeeded.
But what are those \emph{attack-stopping} invariants? There is no ground truth, since the attack went through and stole real money.
We use manual analysis to fill this gap.

\begin{table}
\centering
\caption{Five highest-loss incidents in \dataset, ordered by normalized USD loss.}
\label{tab:dataset-top5}
\begin{tabular}{@{}llll@{}}
\toprule
\textbf{Date} & \textbf{Protocol} & \textbf{Loss} & \textbf{Root Cause} \\
\midrule
Apr 2018 & BeautyChain (BEC)     & \textasciitilde\$900M  & Overflow \\
Apr 2018 & SmartMesh (SMT)       & \textasciitilde\$140M      & Overflow \\
Jun 2024 & WIFCOIN               & \textasciitilde\$13.2M & Logic Flaw \\
Feb 2021 & Yearn yDAI            & \textasciitilde\$11M & Slippage \\
Sep 2024 & OnyxDAO               & \textasciitilde\$3.8M & Logic Flaw \\
\bottomrule
\end{tabular}
\end{table}

\textbf{Manual Analysis} We collect the ground-truth program invariants as follows.
For each attacked contract, two authors independently write a program invariant, time-boxed to eight hours per case.
For this, they consult all possible public information incl. post-mortem reports, the victim contract source code, and the \defihacklabs proof-of-concept exploit.
A third author resolves disagreements if any.
The goal is that each written invariant 1) stops the attack 2) preserves legitimate behavior.

The result is \dataset, a dataset of \datasetsize entries.
Each entry is a triple of 1) a documented exploit transaction 2) a source code proof-of-concept that reproduces the attack path 3) a human-authored invariant.
To illustrate that the dataset only involves past incidents where cybercriminals stole real money, Table~\ref{tab:dataset-top5} lists the five highest-loss incidents in our curated dataset.

\subsection{Verification Procedure}
\label{sec:verification}

When one adds an invariant to an existing smart contract, we call the resulting smart contract a ``hardened contract''.
A hardened contract must satisfy two properties.
It must be \emph{secure} against the exploit that drained the original victim, and it must stay \emph{functional} for legitimate use.

\paragraph{Security.}
We test security by passing the hardened contract through two independent tests: the \dataset proof-of-concept and the original malicious on-chain transaction(s).

The proof-of-concept validates that the invariant breaks the exploit logic per the community post-mortem and the reproduced attack path.

The attack transaction validates that the invariant breaks the attack with the exact calldata, gas, and call path exercised in the real attack. Here, the semantics of blockchain is key: the blockchain gives full guarantees that the on-chain state can be perfectly reproduced at the time of the attack (same block, same transactions before the attack transaction). This is implemented with the mature technique of forking mainnet.
To sum up, we verify that the attack transaction reverts on the hardened contract.

In the following, the hardened contract is considered secure if and only if both replays fail. We note that a side-effect implication of this protocol is to verify that the proof-of-concept exploit is faithful to the attack.

\paragraph{Behavioral Preservation.}
In Solidity, an invariant in a require statement can have side-effects if it passes.
In other words, the added invariant in the hardened contract may impact the behavior for legitimate cases.
Yet, we want invariants that block the attack without changing how the contract behaves for legitimate callers.
We need to check for behavioral preservation.

To assess behavioral preservation, we collect every historical transaction that 1) happens before the attack transaction and 2) reaches the function(s) containing the program invariant. 
For each collected historical transaction, we then check its outcome against the hardened contract instead of the vulnerable one.


\subsection{Verification Infrastructure}
\label{sec:pondereplay}
As stated in \autoref{sec:verification}, our entire methodology is based on replaying transactions. The replay is sophisticated; it must support replaying 1) the original bytecode 2) the recompiled source code and 3) any modification of the contract, in particular the hardened contract.

To the best of our knowledge, there is no tooling for this.
The existing solutions cover only half the problem: \texttt{eth\_call} state overrides can substitute
bytecode but lack sub-block state granularity, forcing a choice between the pre-execution state at $N{-}1$
(missing same-block prior transactions) and the post-execution state at $N$; \texttt{cast run} replays a
historical transaction faithfully but provides no mechanism to substitute the bytecode at any address; 
Foundry fork tests with \texttt{vm.etch} can inject arbitrary bytecode but require a hand-reconstructed scenario rather than the real transaction's context.
No existing tool supports historical replay verification.

We build \pondereplay to fill this gap.
\pondereplay can replay a transaction against a different version of the contract bytecode deployed at a certain target address.
It answers the counterfactual question Section~\ref{sec:verification} needs: what would this real transaction have done if the contract had been hardened?

\subsubsection{Replay Setup}
For a transaction in block $N$, \pondereplay forks Ethereum mainnet at block $N{-}1$.
It injects the new bytecode at the target contract's on-chain address.
If the target transaction is not the first transaction in block $N$, \pondereplay replays the previous transactions in the block before executing the target transaction.
This reconstructs the state immediately before the transaction under study.

The injected bytecode must be compatible with the deployed contract.
It must be compiled with the same EVM version as the original; otherwise, some opcodes are not yet supported in blocks before the intended EVM upgrade.


\definecolor{cmpfill}{HTML}{F1F4F9}
\definecolor{cmpedge}{HTML}{3B5C8F}

\begin{figure}[t]
  \centering
  \resizebox{\linewidth}{!}{%
  \begin{tikzpicture}[
    font=\footnotesize,
    trace/.style  = {rectangle, rounded corners=2pt, draw=black!60, fill=white, thick,
                     align=left, inner sep=4pt, minimum height=12mm, text width=42mm},
    cmp/.style    = {rectangle, rounded corners=3pt, draw=cmpedge!70, fill=cmpfill,
                     line width=1pt, align=center, inner sep=3pt, minimum height=13mm, text width=30mm},
    legend/.style = {rectangle, rounded corners=2pt, draw=black!30, fill=black!3,
                     align=center, inner sep=3pt, font=\scriptsize},
    cmparr/.style = {-{Stealth[length=2.2mm,width=1.8mm]}, thick, draw=cmpedge!80}
  ]

  \node[trace] (O) at (-20mm,  0mm)
    {\textbf{O}~~\emph{original}\\[1pt]
     {\scriptsize literal on-chain data,\\[-2pt] canonical chain, block $N$}};
  \node[trace] (R) at (-20mm,-19mm)
    {\textbf{R}~~\emph{recompiled}\\[1pt]
     {\scriptsize unmodified source code,\\[-2pt] mainnet fork at $N{-}1$}};
  \node[trace] (H) at (-20mm,-38mm)
    {\textbf{H}~~\emph{hardened}\\[1pt]
     {\scriptsize invariant installed,\\[-2pt] mainnet fork at $N{-}1$}};

  \node[cmp] (c1) at (24mm, -9.5mm)
    {\textbf{Reproduction}\\ $\mathbf{R} \overset{?}{\equiv} \mathbf{O}$\\[1pt]
     {\scriptsize $\neq$ $\Rightarrow$ \emph{inconclusive}}};
  \node[cmp] (c2) at (24mm,-28.5mm)
    {\textbf{Preservation}\\ $\mathbf{H} \overset{?}{\equiv} \mathbf{R}$\\[1pt]
     {\scriptsize $\neq$ $\Rightarrow$ \emph{not preserved}}};

  \draw[cmparr] (O.east) -- (c1.north west);
  \draw[cmparr] (R.east) -- (c1.south west);
  \draw[cmparr] (R.east) -- (c2.north west);
  \draw[cmparr] (H.east) -- (c2.south west);

  \node[legend] at (0,-49mm) (leg)
    {Both checks compare: transaction status \,\textbullet\,
     subcall outcomes \,\textbullet\, storage diff ($\epsilon$-threshold)};

  \end{tikzpicture}%
  }
  \caption{\pondereplay{} analyzes replay fidelity.
    Each transaction replay yields three execution traces:
    the literal on-chain record~(\textbf{O}), a recompiled replay of the unmodified
    contract~(\textbf{R}), and a replay of the hardened contract~(\textbf{H}).
    The \emph{reproduction check} ($\mathbf{R}\overset{?}{\equiv}\mathbf{O}$) validates replay
    infrastructure; failure renders the transaction inconclusive.
    The \emph{preservation check} ($\mathbf{H}\overset{?}{\equiv}\mathbf{R}$) isolates the
    invariant's effect; any divergence is causally attributable to the injected guard.}
  \label{fig:pondereplay}
\end{figure}
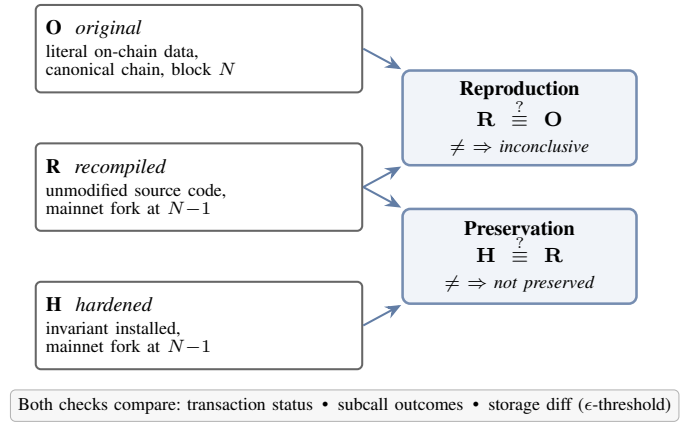

\subsubsection{Replay Fidelity}
\label{sec:replay-fidelity}

A \emph{replay} means re-executing one transaction's calldata against another version of the target contract. A replay produces three traces, illustrated in Figure~\ref{fig:pondereplay}.
The \emph{original} trace (\textbf{O}) is the literal on-chain data.
The \emph{recompiled} trace (\textbf{R}) is obtained by re-running the transaction after injecting a recompiled version of the unmodified contract, using the same source and deployment configuration as the on-chain version of the contract; it is used for sanity checking.
The \emph{hardened} trace
(\textbf{H}) is obtained by re-running the transaction after injecting the bytecode of the hardened contract.

Then, one needs to compare the traces.
Transaction status alone is insufficient: the side effects of a passing transaction might be vastly different or a revert can occur deep in a subcall while the top-level call succeeds, so a status-only check would miss the divergence\footnote{In Solidity, subcall reverts do not revert the top-level transaction -- alas.}.

To assess the fidelity of the replay, \pondereplay verifies three strong criteria: transaction status fidelity, subcall fidelity, and storage fidelity.
Transaction status fidelity is a binary outcome, naturally supported by the EVM.
Subcall fidelity means that the set of all contract calls and their status in the transaction is the same.

Storage fidelity means that the side effects on chain are identical enough, per numerical semantics. In this case, any divergence is considered as a numerical divergence, and we check that the storage value difference is lower than an arbitrary threshold  $\epsilon$.
For storage check, \pondereplay uses a \texttt{prestateTracer} diff to record which slots changed and by how much.

Together the three criteria cover the full observable footprint of a transaction: whether it succeeded, what internal calls change along the way, and what state it ultimately committed.

\subsubsection{Replay Checks}

\textbf{The reproduction check} (\textbf{O} vs.\ \textbf{R}) asks whether the replay infrastructure faithfully reconstructs the on-chain execution.
It is necessary for logically attributing behavioral differences to the invariant.
If the recompiled replay cannot match the original on-chain outcome on all three axes, any conclusion later drawn from \textbf{H} on this transaction would be \emph{inconclusive}: the discrepancy could stem from replay error rather than the invariant.

\textbf{The preservation check} (\textbf{R} vs.\ \textbf{H}) isolates the sole variable that differs between the two executions: the injected invariant.
Because \textbf{R} and \textbf{H} share the same block fork, the same pre-transaction state, and the same calldata, any divergence on the three axes is exclusively attributable to the invariant.
This gives a clean causal guarantee: a divergence is the invariant's effect, not an artifact of the replay environment.

\subsection{Research Questions}
We organize the study around four research questions:
\begin{itemize}
    \item \textbf{RQ1.} \rqone
    \item \textbf{RQ2.} \rqtwo
    \item \textbf{RQ3.} \rqthree
    \item \textbf{RQ4.} \rqfour
\end{itemize}

\subsubsection{RQ1: \rqone} \label{rq1-methodology}

RQ1 studies whether the manually written invariants stop the considered real-world attacks.
For each hardened contract, we replay the two attack views from the security check (Section~\ref{sec:verification}): the \dataset proof-of-concept and the on-chain attack transaction.
The invariant is accepted as effective when both fail at the vulnerable step.

\textbf{Invariant taxonomy.}
We also want deep qualitative knowledge about the kinds of invariants \dataset contains.
We define a unified taxonomy of eight invariant families, reconciling the properties from Bernardi et al.~\cite{bernardi2020wip}, the templates from Trace2Inv~\cite{chen_demystifying_2024}, and the \defihacklabs root cause metadata:
1)~access control \& authorization;
2)~ledger, supply \& conservation;
3)~business logic \& protocol rules;
4)~price, oracle \& slippage;
5)~input validation \& preconditions;
6)~arithmetic \& precision safety;
7)~external call \& interaction safety, incl.\ reentrancy \& execution ordering;
8)~initialization, storage \& configuration.
We manually assign each invariant to one family based on the protocol specification and the intent of the hardened function.

\begin{table*}
  \centering
  \caption{RQ1 attack-stopping invariants in \dataset. \emph{Loss}: as reported by \defihacklabs; losses denominated in ETH are converted to USD at the ETH/USD market price at the attack transaction's block timestamp. \emph{Taxonomy}: family from the unified taxonomy (Section~\ref{rq1-methodology}) assigned from the invariant's semantics. \emph{Hardened fn.}: function(s) where the invariant was inserted (long names fold or abbreviate with \ldots). \emph{Invariant}: source code of the invariant. \emph{Security}: the core RQ1 results, whether the contract hardened with the added invariant blocks both the \defihacklabs PoC and the malicious on-chain transaction (Tx).}
  \label{tab:rq1results}
  \vspace{-4pt}
  \adjustbox{max size={\textwidth}{0.48\textheight},center}{%
  \scriptsize
  \renewcommand{\rqcode}[1]{{\ttfamily\scriptsize\seqsplit{#1}}}%
  \begin{tblr}{
    colspec = {l l c l l Q[l,0.5\linewidth] c c},
    colsep = 2pt,
    rowsep = 0pt,
  }
  \toprule
  & & & & & & \SetCell[c=2]{c} Security \\
  \cmidrule[lr]{7-8}
  Date & Case & Loss & Taxonomy & Hardened fn. & Invariant & PoC & Tx \\
  \midrule
  2018-04 & BEC             & \$900M            & Arithmetic        & \rqfn{batchTransfer}                                   & \rqcode{require(\_value <= uint256(-1) / cnt)}                                                         & \cmark & \cmark \\
  2018-04 & SmartMesh       & \$140M            & Arithmetic        & \rqfn{transferProxy}                                   & \rqcode{require(total >= \_feeSmt \&\& total >= \_value)}                                              & \cmark & \cmark \\
  2020-08 & Opyn            & \$370K            & Ledger/supply     & \rqfn{exercise}                                        & \rqcode{require(msg.value == underlyingRequiredToExercise(oTokensToExercise))}                         & \cmark & \cmark \\
  2021-02 & Yearn yDAI      & \$11M             & Access control      & \rqfn{earn}                                            & \rqcode{require(msg.sender == governance, "only governance can call earn()")}                        & \cmark & \cmark \\
  2021-09 & Nimbus          & \$4.9K          & Price/oracle      & \rqfn{swap}                                            & \rqcode{require(...mul(\_reserve0).mul(\_reserve1).mul(10\_000**2), "Nimbus: K")}                     & \cmark & \cmark \\
  2022-01 & Anyswap         & \$1.4M            & Interaction safety & \rqfnshort{anySwap...Permit}                                & \rqcode{require(returnData.length > 0, "Router: permit function missing for relayer");}                                & \cmark & \cmark \\
  2022-02 & TecraSpace      & \$63K             & Ledger/supply      & \rqfn{burnFrom}                                        & \rqcode{require(\_allowances[from][msg.sender] >= amount)}             & \cmark & \cmark \\
  2022-06 & Inverse Finance & \$1.26M           & Price/oracle      & \rqfn{latestAnswer}                                    & \rqcode{require(crvLPTokenPrice >= lower \&\& crvLPTokenPrice <= upper, "LP price deviates")}          & \cmark & \cmark \\
  2022-09 & BadGuysbyRPF    & \$4.1K & Ledger/supply      & \rqfn{WhiteListMint}                                   & \rqcode{require(\_numberMinted(msg.sender) + chosenAmount <= 1, "Mint limit exceeded");}                                      & \cmark & \cmark \\
  2022-10 & N00d            & \$29K             & Interaction safety & \rqfn{enter}                                           & \rqcode{modifier reentrancy\_lock(){...}}                                            & \cmark & \cmark \\
  2022-10 & Uerii           & \$2.4K            & Ledger/supply     & \rqfn{\_mint}                                          & \rqcode{require(totalSupply() + amount <= CAP)};       & \cmark & \cmark \\
  2022-12 & JAY             & \$18K         & Interaction safety & \rqfn{buyJay}, \rqfn{sell}               & \rqcode{modifier reentrancy\_lock(){...}}                            & \cmark & \cmark \\
  2023-01 & QTN             & \$3.1K             & Access control      & \rqfn{\_transfer}                                      & \rqcode{require(msg.sender == address(uniswapV2Router), "rebase denied")}                              & \cmark & \cmark \\
  2023-05 & ERC20TokenBank  & \$111K            & Price/oracle      & \rqfn{doExchange}                                      & \rqcode{require(namount >= (camount * 995) / 1000, "slippage too high")}                               & \cmark & \cmark \\
  2023-06 & VINU            & \$6K              & Input validation  & \rqfnshort{addLiq...ETH}                                 & \rqcode{require(extcodesize(devaddr) == 0, "devaddr not a contract")}                                 & \cmark & \cmark \\
  2023-08 & Uwerx           & \$327K           & Init/config       & \rqfn{\_transfer}                                      & \rqcode{require(uniswapPoolAddress!=address(0x1),"uniswapPoolAddress not initialized");}                           & \cmark & \cmark \\
  2023-09 & JumpFarm        & \$3.9K           & Ledger/supply     & \rqfn{unstake}                                         & \rqcode{require(TOKEN.balanceOf(this) <= balanceBefore, "rebase mid-unstake")}                         & \cmark & \cmark \\
  2023-09 & Unicly NFT      & \$1.3K           & Interaction safety & \rqfn{deposit}, \rqfn{withdraw}          & \rqcode{modifier reentrancy\_lock(){...}}                         & \cmark & \cmark \\
  2023-10 & pSeudoEth       & \$2.3K            & Price/oracle      & \rqfn{skim}                                            & \rqcode{require(balance0 - reserve0 <= reserve0/10); require(balance1 - reserve1 <= reserve1/10)}    & \cmark & \cmark \\
  2023-11 & grok            & \$55K            & Price/oracle      & \rqfnshort{swap...ForEth}                                              & \rqcode{require(swapAmount <= taxAmount, "Autoswap exceeds sell tax");}                 & \cmark & \cmark \\
  2024-04 & HoppyFrog       & \$979           & Price/oracle      & \rqfn{\_transfer}, \rqfnshort{swap...ForEth}                         & \rqcode{require(swapAmount <= maxSwapForSell, "Autoswap exceeds tax-scaled cap");}                & \cmark & \cmark \\
  2024-06 & APEMAGA         & \$30K             & Access control      & \rqfn{family}                                          & \rqcode{require(msg.sender == account, "Unauthorized caller")}                                         & \cmark & \cmark \\
  2024-06 & JokInTheBox     & \$34K           & Business logic     & \rqfn{unstake}                                         & \rqcode{require(!currentStake.unstaked, "Stake already unstaked")}                                    & \cmark & \cmark \\
  2024-06 & WIFCOIN         & \$13.19M          & Business logic     & \rqfn{claimEarned}                                     & \rqcode{require(\_earned > totalRewardsPerWalletPerPlan[...], "Rewards already claimed")}              & \cmark & \cmark \\
  2024-08 & OMPx            & \$11K          & Price/oracle      & \rqfn{purchase}, \rqfn{buyBack}          & \rqcode{modifier enforceCooldown{...}}                  & \cmark & \cmark \\
  2024-09 & Bedrock DeFi    & \$2M              & Price/oracle    & \rqfn{mint}                                            & \rqcode{require(uniBTCAmount * 1e10 < msg.value, "Exchange rate unsafe")}                              & \cmark & \cmark \\
  2024-09 & OnyxDAO         & \$3.8M            & Ledger/supply     & \rqfnshort{liquidate...Repay}                                      & \rqcode{require(repayAmount == borrowedAmount)}                                    & \cmark & \cmark \\
  2026-03 & AlkemiEarn      & \$89K         & Access control      & \rqfn{liquidateBorrow}                                 & \rqcode{require(msg.sender != targetAccount, "self-liquidation not possible")}                         & \cmark & \cmark \\
  \midrule
 \textbf{Total 28} & \textbf{Total loss} &\textbf{\$1.07B} & &  & & \textbf{28} & \textbf{28} \\
  \bottomrule
  \end{tblr}%
  }
  \end{table*}

\subsubsection{RQ2: \rqtwo} \label{rq2-methodology}

RQ1 shows that \dataset program invariants stop attacks; RQ2 asks what that protection costs in legitimate usability.
For each hardened contract in \dataset, we replay the contract's historical transaction history through the functional-preservation check (Section~\ref{sec:verification}).
We count how many transactions are preserved and we record the divergence, if any, between the  transaction trace of the recompiled original contract \textbf{R} and the hardened version \textbf{H}.

A preserved transaction means that the hardened contract reproduces the original outcome per the three strong properties verified in the verification infrastructure, see \autoref{sec:replay-fidelity}.

For every transaction that diverges, we manually inspect its call trace and storage diff to understand its root cause.
Once the root cause is understood, we assess what the non-preserved behaviour means for those particular transactions, and we record this assessment as the \textit{behavioral change reason}.

\subsubsection{RQ3: \rqthree} \label{rq3-methodology}

RQ3 studies the transactions that are inconclusive per our replay algorithm (neither preserved nor non-preserved, $\textbf{O} \neq \textbf{R} $).
For each inconclusive transaction, we analyze the original on-chain trace and the recompiled replay trace.
This enables us to understand what difference in the trace causes this replay to be inconclusive.

As an additional check on replay infrastructure, we also test whether the recompiled contract and the hardened one still agree despite the inconclusive verification: agreement here shows that the invariant's effect is the same and that inconclusiveness stems from the replay infrastructure.

\subsubsection{RQ4: \rqfour} \label{rq4-methodology}
RQ4 evaluates how well the state-of-the-art of invariant inference is able to identify the  ground truth invariants.
We are able to run end-to-end two state-of-the-art invariant inference tools, InvCon~\cite{liu_invcon_2022} and InvCon+~\cite{invconplus}, and one invariant generator, FLAMES~\cite{eshghie_flames_2025}. We compare their outputs against the validated invariants from \dataset.

Reapplying our verification protocol is not suitable for this task because the tools do not deliver specific-enough instructions for hardening the contract.
Instead, we rely on checking whether the ground-truth invariant appears in the tool output.

\section{Experimental Results}
\label{sec:results}

\subsection{RQ1: Invariant Effectiveness}

We check whether the invariants make the considered contracts secure.
Table~\ref{tab:rq1results} shows the results.
All \datasetsize ground-truth invariants block: 1) the proof-of-concept exploit 2) the actual attack transaction.
In both cases, the transaction reverts, it reverts at the expected vulnerable step, and it reverts with the revert message written in the invariant.
This perfect result 28/28 (100\%) is a consequence of our ground truth protocol.
There are 25 attacks for which we did not succeed in writing the securing invariant. This is because of our limited expertise of the target protocol and the time-bound budget.

To make the ground truth concrete, we walk through one representative entry of \dataset, from the vulnerability to the attack-stopping invariant.

\textbf{Case Study: BEC} (April 2018, Loss: 900 M USD)
BEC is an ERC-20 token. Its \texttt{batchTransfer()} function pays multiple recipients the same amount in one call by computing \texttt{amount = cnt * value}, checking the sender can afford it, then crediting each recipient. 
However, the multiplication had no safety check. 
The attacker sent the transaction with 2 receivers and a value of \(2^{255}\). 
Multiplying those together (\(2 \times 2^{255}\)) caused an overflow, so the balance check passed even though they had no tokens. 
The inserted program invariant catches such scenario.

\begin{lstlisting}[style=diff, caption={BEC \texttt{batchTransfer()}: attack-stopping invariant.},float]
function batchTransfer( address[] _receivers, uint256 _value ) public whenNotPaused returns (bool) {
   ...
@\diffaddmark{+    require(\_value <= uint256(-1) / cnt);}@
   balances[msg.sender] = balances[msg.sender].sub(amount);
   for (uint256 i = 0; i < cnt; i++) {
       balances[_receivers[i]] = balances[_receivers[i]].add(_value);
        ...
    }
    return true;
}
\end{lstlisting}

\textbf{Taxonomy.}
We analyze our ground truth invariants.
Table~\ref{tab:rq1results} assigns each invariant to a family from the unified taxonomy defined in Section~\ref{rq1-methodology}.
Price oracle and slippage are the largest family, with 8/28 cases.
Ledger, supply, and conservation checks follow with 6 cases.
Access control and authorization account for 4 cases, tied with external call and interaction safety, three of which are reentrancy locks.
Business logic guards and arithmetic safety cover 2 cases each; input validation and initialization checks cover 1 case each.
All eight families are represented in \dataset, confirming that real-world attacks are stopped by a broad spectrum of properties rather than a single dominant pattern.

\begin{answerbox}{Answer to RQ1 } \textbf{\rqone}

Every invariant in \dataset blocks the attack: all \datasetsize hardened contracts cause both the \defihacklabs proof-of-concept and the malicious on-chain transaction to fail at the vulnerable step.
The eight-family taxonomy shows that no single class of guard dominates.
A single \texttt{require} invariant, correctly specifying the protocol's intended security property, is sufficient to abort past real-world attacks totaling more than one billion dollars of theft.
\end{answerbox}

\subsection{RQ2: Behavioral Preservation}

We evaluate the effect on the hardened version of the contract on historical transactions. 
We use \pondereplay to do so, see \autoref{sec:pondereplay}. 
Overall, we replay \totalnumreplayed transactions with our fidelity checks.

\emph{Preserved transactions.} $106{,}815$ ($98.3\%$ of total replayed) transactions pass the preservation check, indicating equivalent behaviour between the recompiled and hardened contracts; this is presented in column Pres. of \autoref{tab:rq2results}.

\emph{Not-preserved transactions.} We observe $1{,}526$ txs ($1.4\%$ of total replayed) with a divergence when replayed under the hardened bytecode (Table~\ref{tab:rq2results}, column Not pres.).

\emph{Divergence Kind} In $1{,}424$ cases ($93\%$), the introduced program invariant reverts within the contract directly.
In $43$ cases, the introduced program invariant reverts in a nested subcall.
In the remaining $59$ cases, the preservation check fails because of \emph{storage drift}, which violates storage fidelity.

\subsubsection{Manual Analysis}
We manually analyze all cases in order to identify the reason for non-preservation.
This reason is documented in Table~\ref{tab:rq2_notpreserved} as column \emph{behavioral change reason}.

\paragraph{Past Exploit.}
A past transaction is classified as \emph{exploit} when it reproduces the same vulnerability call pattern as the documented attack.
The entry point, and triggering condition match what the proof-of-concept or on-chain malicious transaction used to perform the attack.
The transaction reverting in \textbf{H} is intentional and demonstrates soundness.
APEMAGA illustrates this: all $10$ not-preserved calls invoke \texttt{family()} on the same pair, at block $20175377$, \texttt{0xcfbc3\dots} (blocks $20175128$--$20175234$) and \texttt{0xdbab5d\dots} (blocks $20175265$--$20175350$).
The past transactions exercise the same unauthenticated-burn path against the same pair, inflating the token price and ultimately swapping the tokens.
Each call reverts with our ground truth invariant (\texttt{Unauthorized: caller is not the token owner}), the same guard that stops the documented attack. Hence, the invariant is correct and the behavioral change is sound.


\paragraph{Misusage.}
A transaction is classified as \emph{misusage}, if it calls a function in a way the protocol design never envisioned or authorized, while not being malicious.
The caller misuses an accounting rule that the vulnerable implementation left open, for example invoking a privileged internal operation through an unauthenticated wrapper.
The protective invariant restores the intended protocol rule; consequently the replayed transaction correctly reverts.
Nimbus is representative: $449$ of the $466$ not-preserved calls invoke \texttt{swap()} and revert with the constant-product invariant that also stops the attack.
These callers are not attackers; they are ordinary swappers whose trade pushes the pool's reserves past the protocol rule, a path the vulnerable pool tolerated but the protocol design never intended to expose.

\paragraph{Whitehat operation.}
This label covers cases where the transaction does exercise the same code path as the attack, yet was done by a whitehat hacker.
We are able to differentiate exploits from whitehat operations, thanks to Etherscan labels.
Whitehat transactions differ in intent, funds are recovered rather than stolen, but are technically equivalent to an exploit, hence are correctly reverted by the protective invariant.
The Anyswap case is as such.
Etherscan labels \texttt{0x636044\dots} as the \emph{Whitehat}: it calls \texttt{anySwapOutUnderlyingWithPermit} on pools the original attacker had not yet drained, recovering funds before a second attacker could.

%
\begin{table}
\centering
\caption{RQ2 functional preservation on \dataset. \emph{Collected}: pre-attack historical transactions scoped to the hardened function. \emph{Pres.}: behavior preserved under the invariant. \emph{Not pres.}: behavioral divergence under the invariant. \emph{Incon.}: inconclusive, the recompiled replay does not reproduce the on-chain outcome (analyzed in RQ3). Cases marked ``---'' have no reachable pre-attack history for the hardened function.}
\label{tab:rq2results}
\begin{tblr}{
  colspec = {l l r r r r},
  colsep = 2pt,
  rowsep = 0pt,
  vline{4}={solid},
}
\toprule
Date & Case &
  \shortstack{Collected} &
  Pres. & Not pres. & Incon. \\
\midrule
2018-04 & BEC             & 43,952 & 43,951 & 0   & 1   \\
2018-04 & SmartMesh       & 295    & 295    & 0   & 0   \\
2020-08 & Opyn            & 6      & 3      & 3   & 0   \\
2021-02 & Yearn yDAI      & 93     & 4      & 88  & 1   \\
2021-09 & Nimbus          & 862    & 391    & 466 & 5   \\
2022-01 & Anyswap         & 86     & 9      & 77  & 0   \\
2022-02 & TecraSpace      & ---    & ---    & --- & --- \\
2022-06 & Inverse Finance & 49     & 7      & 41  & 1   \\
2022-09 & BadGuysbyRPF    & 769    & 763    & 0   & 6   \\
2022-10 & N00d            & 37     & 37     & 0   & 0   \\
2022-10 & Uerii           & ---    & ---    & --- & --- \\
2022-12 & JAY             & 712    & 711    & 0   & 1   \\
2023-01 & QTN             & 1,406  & 623    & 782 & 1   \\
2023-05 & ERC20TokenBank  & 12     & 10     & 0   & 2   \\
2023-06 & VINU            & 1      & 1      & 0   & 0   \\
2023-08 & Uwerx           & 120    & 120    & 0   & 0   \\
2023-09 & JumpFarm        & 520    & 506    & 0   & 14  \\
2023-09 & Unicly NFT      & 3      & 3      & 0   & 0   \\
2023-10 & pSeudoEth       & ---    & ---    & --- & --- \\
2023-11 & grok            & 56,847 & 56,582 & 4   & 261 \\
2024-04 & HoppyFrog       & 1,949  & 1,895  & 54  & 0   \\
2024-06 & APEMAGA         & 11     & 1      & 10  & 0   \\
2024-06 & JokInTheBox     & 8      & 8      & 0   & 0   \\
2024-06 & WIFCOIN         & ---    & ---    & --- & --- \\
2024-08 & OMPx            & 842    & 839    & 1   & 2   \\
2024-09 & Bedrock DeFi    & ---    & ---    & --- & --- \\
2024-09 & OnyxDAO         & ---    & ---    & --- & --- \\
2026-03 & AlkemiEarn      & 57     & 56     & 0   & 1   \\
\midrule
& \textbf{Total (28/28)} & \textbf{108,637} & \textbf{106,815} & \textbf{1,526} & \textbf{296} \\
\bottomrule
\end{tblr}%
\end{table}

%
\begin{table}
\centering
\caption{Not-preserved historical replays per case. \emph{Not pres.}: transactions where the hardened trace \textbf{H} diverges from the recompiled trace \textbf{R}. \emph{Trace inv.}, \emph{Sub.\ inv.}, and \emph{Tx drift} classify the mechanical cause of divergence: the invariant reverts the top-level call, the invariant reverts inside a subcall, or storage drift. \emph{Behavioural change reason}: case-level root cause from manual analysis.}
\label{tab:rq2_notpreserved}
\vspace{-2pt}
\scriptsize
\begin{tblr}{
  colspec = {llrrrrQ[l,1.7cm]},
  colsep = 2pt,
  rowsep = 0pt,
  vline{4,7}={solid},
}
\toprule
Date & Case &
  \shortstack{Not\\pres.} &
  \shortstack{Trace\\inv.} &
  \shortstack{Sub.\\inv.} &
  \shortstack{Tx\\drift} &
  \shortstack{Behavioural \\ Change Reason} \\
\midrule
20-08 & Opyn             &    3 &    3 &    0 &    0 &  Past Exploit\\
21-02 & Yearn yDAI       &   88 &   88 &    0 &    0 &  Over-restrictive Invariant \\
21-09 & Nimbus           &  466 &  449 &    5 &   12 &  Misusage\\
22-01 & Anyswap          &   77 &   67 &    8 &    2 & Past Exploit \& Whitehat Operation   \\
22-06 & Inverse Finance  &   41 &   33 &    8 &    0 &  Tight Numerical Envelope \\
23-01 & QTN              &  782 &  764 &    0 &   18 & Over-restrictive Invariant  \\
23-11 & grok             &    4 &    1 &    2 &    1 &  Tight Numerical Envelope \\
24-04 & HoppyFrog        &   54 &    8 &   20 &   26 &  Tight Numerical Envelope \\
24-06 & APEMAGA          &   10 &   10 &    0 &    0 &  Past Exploit\\
24-08 & OMPx             &    1 &    1 &    0 &    0 &  Misusage\\
\midrule
& \textbf{Total} & \textbf{1,526} & \textbf{1,424} & \textbf{43} & \textbf{59} &  \\
\bottomrule
\end{tblr}%

\end{table}

\paragraph{Over-restrictive invariant.}
We identify some cases where past transactions revert because the protective invariant is overrestrictive.
In this case, the invariant expresses the right security property but its predicate is narrower than the protocol's actual intended interface.
This means that all reverted transactions were done by honest users; those transactions now fail under $H$ because of the over-restrictive invariant, not because they exploit the vulnerability.
Yearn yDAI is the clearest case: all $88$ not-preserved calls invoke \texttt{earn()} from the same externally owned account, a keeper bot that harvests the vault on a schedule, and each reverts with \texttt{only governance can call earn()}.
The perfect invariant should handle protocol access rules for keeper bots.

\paragraph{Tight numerical envelope.}
There is a particularly interesting variation of over-restrictive invariants, which we call ``tight numerical envelope''.
This category covers invariants whose guard is a numeric bound---a price band, a swap tax, a deviation threshold---rather than an access-control or structural rule.
When the past transaction reverts under replay, this is because the numeric bounds form an envelope too tight for part of the historical traffic.
Reverted transactions under replay are then calibration failures: the invariant would preserve all past transactions with wider or context-sensitive bounds.
Inverse Finance shows this pattern: $33$ of its $41$ not-preserved calls to \texttt{latestAnswer()} revert with \texttt{LP price deviates from Curve VP}, the $\pm10\%$ band the invariant places around the Curve virtual price.
Yet these calls are legitimate oracle reads made during ordinary price volatility, not attacks. 
A wider, volatility-aware, invariant would preserve these reads while still rejecting the flash-loan price spike stopped by the invariant.

\begin{answerbox}{Answer to RQ2 } \textbf{\rqtwo}
The \dataset invariants preserve behavior on $98.3\%$ of replayed historical transactions. Every divergence is causally traced to the added invariant, not to a replay artifact.
Divergences span $10$ cases; some reject legitimate calls through over-restrictive invariants or numeric bounds set too tight, while the rest correctly reject exploits, misuses, or whitehat-rescue transactions.
\end{answerbox}

\subsection{RQ3: Inconclusive Replay}

We now manually analyze the $296$ \textit{inconclusive} replay verifications (column Incon.\ of Table~\ref{tab:rq2results}), which span twelve cases; Table~\ref{tab:rq3inconclusive} breaks them down.
We identify three mechanical causes: \emph{original OOG}, when the transaction ran out of gas under its original gas limit but the replay was allocated more gas and therefore succeeded; \emph{subcall OOG}, when the mismatch is confined to an out-of-gas revert inside a subcall on the original trace rather than the top-level outcome; and \emph{accumulated drift}, when state written by earlier, co-located transactions in the same block differs between real on-chain and the replayed fork.

\textbf{Original OOG} (173 cases). The most frequent cause is a gas limit mismatch between the chain and the replay. On-chain, a transaction is submitted with a caller-specified gas limit; if execution exhausts that budget, the EVM reverts the entire call with an out-of-gas error. \pondereplay, by design, does not cap the replay exactly at the original gas limit: it allocates sufficient gas to be able to execute the invariant. As a result, a transaction that ran out of gas on-chain succeeds in the recompiled replay R. The reproduction check O vs. R then fails, making the transaction inconclusive. This is a deliberate trade-off: enforcing the original gas cap would require exact gas accounting across a fork, which is unreliable.

\textbf{Subcall OOG} (12 cases). A subtler variant occurs when the top-level call succeeds both on-chain and in replay, but an internal subcall exhausts gas and reverts differently in the two executions. In these cases, the transaction status matches between O and R, but the failed-subcall axis diverges: a nested call reverts on-chain but succeeds in the replay, or vice versa. Because the subcall's revert is swallowed by the caller rather than propagating to the top level, the top-level outcome is unaffected. These 12 cases are confined to contracts with deep call trees where one internal path is close to the gas limit.

\textbf{Accumulated drift} (111 cases). The third cause is state drift introduced by co-located transactions in the same block. \pondereplay reconstructs the pre-state of the target transaction by replaying all preceding transactions in block N in order. When any of those prior transactions behaves differently under recompiled bytecode, for example because its own replay is inconclusive, the block state at the point of the target transaction diverges from what actually existed on-chain, causing the target transaction's outcome to change as a consequence. The grok case is the clearest illustration: grok is a high-traffic DEX router through which dozens of other transactions pass in the same block, and small gas or state differences in those co-located calls compound into a state that no longer matches the real chain, accounting for the majority of the $111$ accumulated drift cases.

Finally, we run the agreement check announced in Section~\ref{rq3-methodology}: for $291$ of the $296$ inconclusive transactions ($98\%$), the hardened trace \textbf{H} still matches the recompiled contract outcome \textbf{R} (last column of Table~\ref{tab:rq3inconclusive}).
This confirms that inconclusiveness stems from the replay infrastructure, not from the invariant.

%
\begin{table}
\centering
\caption{Inconclusive historical replays per case. \emph{Total Txs}: total number of transactions for which \textbf{R} does not reproduce \textbf{O}. \emph{Original OOG}, \emph{Sub.\ OOG}, and \emph{Acc.\ drift} classify the inconclusive txs. \emph{Preserved (H$\equiv$R)}: inconclusive transactions where \textbf{R} and \textbf{H} are nevertheless equivalent, i.e., the invariant does not alter the on-chain outcome.}
\label{tab:rq3inconclusive}
\begin{tblr}{
  colspec = {llrrrrr},
  colsep = 2pt,
  rowsep = 0pt,
  vline{4,7}={solid},
}
\toprule
Date & Case &
  \shortstack{Total\\Txs }&
  \shortstack{Original\\OOG} &
  \shortstack{Sub.\\OOG} &
  \shortstack{Acc.\\drift} &
  \shortstack{Preserved\\(H$\equiv$R)} \\
\midrule
18-04 & BEC             &   1 &   1 &   0 &   0 &   1 \\
21-02 & Yearn yDAI      &   1 &   1 &   0 &   0 &   0 \\
21-09 & Nimbus          &   5 &   2 &   3 &   0 &   3 \\
22-06 & Inverse Finance &   1 &   1 &   0 &   0 &   0 \\
22-09 & BadGuysbyRPF    &   6 &   3 &   0 &   3 &   6 \\
22-12 & JAY             &   1 &   0 &   0 &   1 &   1 \\
23-01 & QTN             &   1 &   1 &   0 &   0 &   0 \\
23-05 & ERC20TokenBank  &   2 &   2 &   0 &   0 &   2 \\
23-09 & JumpFarm        &  14 &  13 &   0 &   1 &  14 \\
23-11 & grok            & 261 & 146 &   9 & 106 & 261 \\
24-08 & OMPx            &   2 &   2 &   0 &   0 &   2 \\
26-03 & AlkemiEarn      &   1 &   1 &   0 &   0 &   1 \\
\midrule
& \textbf{Total} & \textbf{296} & \textbf{173} & \textbf{12} & \textbf{111} & \textbf{291} \\
\bottomrule
\end{tblr}%
\end{table}

\begin{answerbox}{Answer to RQ3 } \textbf{\rqthree}

Inconclusive replays are fully understood and consistent.
The dominant cause is out-of-gas changes, either in the called function ($173$ cases) or in subcalls ($12$ cases): the gas semantics is fully responsible for the behavioral change.
Accumulated drift ($111$ cases) stems from co-located transactions in the same block that alter shared state before the target transaction executes. Our original historical replay verification is sound and its counterfactual analysis works as expected.

\end{answerbox}

\subsection{RQ4: Automatically Generated Invariants}

Table~\ref{tab:rq4-capabilities} summarizes the capability of the three reproducible tools to generate effective invariants.
Across InvCon~\cite{liu_invcon_2022}, InvCon+~\cite{invconplus}, and FLAMES~\cite{eshghie_flames_2025}, only 2 of the \datasetsize ground-truth invariants can be synthesized, both by FLAMES.
These two cases correspond to APEMAGA and Inverse Finance.
All three tools produce candidates for many contracts (resp.\ 13, 24, and 28), but the candidates hardly capture the security-critical property that can block the exploit.

\begin{table}
\centering
\caption{Overall capability of the three reproducible tools on \dataset.
  \emph{Generated}: contracts for which the tool produced at least one invariant for the vulnerable function.
  \emph{Match}: subset of \emph{Generated} whose invariant exactly matches the ground truth.
  \emph{Not generated}: contracts for which the tool produced no invariant for the vulnerable function (crash, empty output, or wrong target).}
\label{tab:rq4-capabilities}
\begin{tblr}{
  colspec = {l c c c},
  row{1} = {font=\bfseries},
}
\toprule
Tool & Generated & Match & {Not generated} \\
\midrule
InvCon~\cite{liu_invcon_2022}     & 13 & 0 & 15 \\
InvCon+~\cite{invconplus} & 24 & 0 & 4  \\
FLAMES~\cite{eshghie_flames_2025} & 28 & 2 & 0  \\
\bottomrule
\end{tblr}
\end{table}

\begin{answerbox}{Answer to RQ4 } \textbf{\rqfour}
State-of-the-art invariant generation tools are not yet capable of protecting against real-world attacks.
Only 3 tools from previous research can be executed in this real-world setup.
Only 2 of the \datasetsize ground-truth invariants are synthesized by at least one of them.
The considered tools produce many candidates, but almost none capture the security-critical property that blocks the exploit.
This calls for more research on invariant synthesis, evaluated on the real-world attack frontline. 
\end{answerbox}

\section{Related Work}
\label{sec:related}

This paper sits at the intersection of several research lines.

\subsection{Invariant Generation}
Invariant generation can be defined as proposing  invariants directly from a program's implementation, without requiring execution traces.
Wang et al.~\cite{canllmreason2023} showed that a language model fine-tuned for invariant prediction infers invariants, at a quality comparable to a dynamic detector supplied with several execution traces.
SmartInv~\cite{wang_smartinv_2024} targets defects whose detection needs semantic reasoning that source code alone does not expose.
It reasons jointly over Solidity source code and natural language artefacts such as comments and documentation.
PropertyGPT~\cite{liu_propertygpt_2025} uses GPT-4 with retrieval-augmented generation to produce formal Certora verification properties.
FLAMES~\cite{eshghie_flames_2025} fine-tunes a language model on real-world invariants extracted from over 500{,}000 verified on-chain contracts in order to generate executable \texttt{require} guards directly from source code, without labels or formal specifications.

These tools only report accuracy against synthetic properties, audit reports, or held-out contracts. 
None measures whether the generated invariants  can actually stop real on-chain exploits.
Our paper closes this gap by studying whether generated invariants actually protect smart contracts. We also note that few of these tools offered replication packages of sufficient quality that we can use in RQ4.

\subsection{Invariant Inference}
Invariant inference infers invariants from execution traces.
The best known invariant inference tool is the Daikon dynamic invariant engine~\cite{daikon2007}.
In blockchain, deployed contracts naturally accumulate rich execution histories, and mining these traces is a fundamentally good idea.

There exist several invariant inference tools for smart contracts.
InvCon~\cite{liu_invcon_2022} adapts the Daikon engine to Ethereum, mining on-chain histories as execution traces in order to flag ERC-20 non-compliance.
InvCon+~\cite{invconplus} pairs dynamic trace mining with a Houdini-based static verifier.
This filtering yields invariants that are formally sound, unlike the merely likely invariants InvCon reports.

Trace2Inv~\cite{chen_demystifying_2024} applies dynamic taint analysis to find which variables influence sensitive operations, then instantiates a catalogue of invariant templates as contract-specific \texttt{require} invariants.
OpenTracer~\cite{Chen2024OpenTracerADAK} is Trace2Inv's trace-analysis engine, released as a standalone dynamic analyzer for invariant generation.

SmartOracle~\cite{su_smartoracle_2025} mines fine-grained invariants that span function calls, token balances, and state changes within and across transactions.
This cross-transaction and cross-contract scope detects violations that single-trace invariants miss.

Every inference tool depends on trace availability: under-exercised functions accumulate too few data points to yield reliable invariants~\cite{chen_demystifying_2024}. By construction, invariant inference tools cannot protect a newly deployed protocol.
None of these works quantifies, as we do, how mined invariants are successful at stopping real-world attacks.

\subsection{Contract Hardening}
Another line of work studies contract hardening.
Sereum~\cite{Rodler2018SereumPEI} protects already-deployed contracts against reentrancy by tracking data flow inside a modified EVM and aborting transactions that violate a fixed reentrancy invariant.
ContractGuard~\cite{Wang2019ContractGuardDEW} introduces a runtime anomaly detection system for Ethereum smart contracts that profiles execution paths and reverts attacks upon detecting abnormal control flows.
\aegis~\cite{Torres2020GISSVAS} proposes a runtime protection mechanism for deployed smart contracts that matches transactions against attack patterns.

Solythesis~\cite{Li2020SecuringSCO} transforms Solidity source to insert runtime checks for developer-specified invariants, and reduces the resulting gas overhead through incremental state updates.
HCC~\cite{Giesen2022HCCALF} is a hardening compiler that automatically injects integrity guards against reentrancy and integer bugs.
HighGuard~\cite{Eshghie2023HighGuardCBBF} detects business-logic violations at runtime by comparing execution against a formal model of intended behavior.

BACKRUNNER~\cite{Shou2024BACKRUNNERMSAZ} mitigates attacks by monitoring pending mempool transactions and rescuing funds before an exploit settles. 
CrossGuard~\cite{Chen2025EnforcingCFK} constrains the permitted sequences of external calls to block unexpected interactions.

Raven~\cite{eshghie_raven_2025} mines defensive patterns from Ethereum by clustering the conditions under which real transactions revert, yielding semantic categories of protective invariants.

\subsection{On-chain Exploit Datasets}

Empirical studies characterize the attacks themselves.
Zhou et al.~\cite{Zhou2020AnEGAV} study real world attacks and their potential mitigations.
Zhang et al.~\cite{Zhang2022CombattingFIAQ} mine real front-running attacks to evaluate detectors.
A recent systematic review attributes over one billion dollars of losses to recurring vulnerability root causes~\cite{Rezaei2025SoKRCAW}.
These works only document what went wrong.

To our knowledge, we are the first to evaluate the extent to which smart contract invariants can protect against real-world attacks.

\section{Conclusion}

In this paper, we studied whether program invariants stop real smart contract exploits and whether automated tools could discover them.
We built \dataset, a benchmark of \datasetsize real Ethereum attacks each paired with a human-authored invariant.
We validated every invariant with a novel replay framework that re-executed \totalnumreplayed historical transactions.
Every invariant blocked its target attack; $98.3\%$ of replayed historical transactions were preserved, and every deviation was traced causally to the invariant rather than to replay noise.
Against this ground truth, state-of-the-art generation tools fell short: FLAMES recovered the correct guard in only $2$ of $28$ cases.
A single correctly placed invariant is sufficient to stop a real exploit; finding that invariant automatically remains an open problem.

 \section*{Acknowledgements}
This work was supported by the WASP program funded by Knut and Alice Wallenberg Foundation, and by the Swedish Foundation for Strategic Research (SSF). Some computation was enabled by resources provided by the National Academic Infrastructure for Supercomputing in Sweden (NAISS).
We thank Dwellir for giving us access to their RPC infrastructure under an academic license.

\bibliographystyle{plain}
\bibliography{main}

\end{document}